\documentclass[12pt,english,floatfix,superscriptaddress,aps,prd,preprint,showkeys]{revtex4}
\usepackage{amsmath}
\usepackage{amssymb}
\usepackage{amsbsy}
\usepackage{amsfonts}
\usepackage{amsopn}
\usepackage{amstext}
\usepackage{graphicx}
\usepackage[english]{babel}
\usepackage{color}
\usepackage{slashed}
\usepackage{esint}
\usepackage{float}
\usepackage{units}
\usepackage{textcomp}
\usepackage{wasysym}

\usepackage{slashed}
\usepackage{hyperref}
\usepackage{graphicx}
\usepackage{amssymb}
\usepackage{amsmath}
\usepackage{color}

\DeclareMathOperator{\sech}{sech}

\begin{document}

\title{Torsion braneworlds in a tensor-vector gravity}

\author{J. E. G. Silva}
\email{euclides.silva@ufca.edu.br}
\affiliation{Universidade Federal do Cariri(UFCA), Av. Tenente Raimundo Rocha, \\ Cidade Universit\'{a}ria, Juazeiro do Norte, Cear\'{a}, CEP 63048-080, Brasil}

\author{L. J. S. Sousa}
\email{luisjose@fisica.ufc.br}
\affiliation{Instituto Federal de Educa\c{c}\~{a}o Ci\^{e}ncia e Tecnologia do Cear\'{a} (IFCE) - Campus de Maracana\'{u} Av. Contorno Norte, 10 - Distrito Industrial, Maracana\'{u}, 61936-000 - Cear\'{a} - Brazil}

\author{W. T. Cruz}
\email{wilami@ifce.edu.br}
\affiliation{Instituto Federal de Educa\c{c}\~ao, Ci\^encia e Tecnologia do Cear\'a (IFCE), Campus Juazeiro do Norte - 63040-540 - Juazeiro do Norte, CE - Brazil}

\author{C.A.S. Almeida}
\email{carlos@fisica.ufc.br}
\affiliation{Universidade Federal do Cear\'a (UFC), Departamento de F\'isica,\\ Campus do Pici, Fortaleza - CE, C.P. 6030, 60455-760 - Brazil.}

\begin{abstract}
We study the properties of gravity and  bulk fields living in a torsion warped braneworld. The torsion is driven by a background vector whose norm provides a source for the bulk cosmological constant. For a vector as the derivative of a scalar field, we find new isotropic thick brane geometry. We analyze the features of bosonic and fermionic fields in this scenario. The background vector provides nonminimal coupling between the fields and the geometry leading to modifications in the Kaluza-Klein states. For a scalar field, the background vector breaks the $Z_2$ symmetry, whereas for the vector gauge field a geometric coupling to the torsion tensor traps the massless mode. The spinor connection is modified by the torsion and a geometric derivative Yukawa-like coupling is proposed. The effects of these new couplings are investigated.
\end{abstract}

\keywords{Modified gravity, braneworld, Lyra geometry, torsion}

\maketitle

\section{Introduction}
In braneworld paradigm our $4D$ world is viewed as a brane embedded in a higher dimensional bulk spacetime. These models provide geometrical solutions for some of greatest problems in physics, such as the the hierarchy problem \cite{rs1}, the origin of dark matter \cite{Arkani1998}, the cosmological constant problem \cite{Chen2000} and the Big Bang singularity \cite{Khoury2001}.

The main idea here is to define the gravity and field dynamics in the bulk and obtain the effective dynamics on the brane by dimensional reduction. Several  mechanisms were proposed for bosonic, fermionic and gravity fields \cite{Kehagias2001, nosso4, casa,blochbrane,Cruz2016}. The dimensional reduction yields the standard model (SM) action on the brane and a tower of massive Kaluza-Klein (KK) states. The KK spectrum signals for a physics beyond the standard model \cite{kkbeyondsm}.

The extra dimensions also allow us to propose non-standard bulk dynamics and seek for their phenomenological effects on the brane. Notably, models involving supersymmetry \cite{susy}, Lorentz violation \citep{lv}, noncommutative geometry \citep{nc}, orbifold internal space \cite{Silva2011} and minetic gravity \cite{minetic} were investigated. 

An important geometric extension in braneworld models is achieved considering a Weyl geometry, wherein the metric compatibility is modified due to a background vector  \citep{weyl,weyl1, weyl2}. This background vector breaks the spacetime integrability which can be restored provided the Weyl vector be the gradient of a scalar function, called the Weyl scalar potential. For a particular choice of the Weyl scalar potential, thick brane solutions were found in this vacuum Weyl spacetime \citep{weyl,weyl1, weyl2,weyl3}. This integrable Weyl braneworld share many properties with the Einstein-dilaton brane models   \cite{Tao2017}. 

Another interesting braneworld extension arises due to the inclusion of a torsional bulk. Torsion breaks the symmetry of the connection while can preserves the metric compatibility \cite{hehl,shapiro}. In string theory, notably the heterotic theory, the skew-symmetric Kalb-Ramond gauge field can be regarded as a torsion source \citep{torsion1,torsion2,kalbramond,Becker}. Thus, the open string spectrum provides bulk extensions of Riemannian geometry. Among the torsion effects on the brane we outline the modified black hole horizon \citep{roldao}, a correction to the brane cosmological constant \citep{roldao1}, $CPT$ violation \citep{ellis} and the split of the brane in $f(T)$ teleparallel gravity \citep{Yang}. Furthermore, the coupling of torsion with standard model fields can yields Lorentz symmetry breaking \cite{kostelecky1}.

In this article we propose a torsional braneworld whose torsion is driven by a background vector, in the so called Lyra geometry \citep{lyra, Sen-1972}. The asymmetric Lyra connection can be recast into a symmetric (compactible) connection by including the torsion terms in the connection coefficients. Likewise the Weyl geometry, the resulting gravitational action can be regarded as a tensor-vector theory \citep{Sen-1971, lyracosmology}. The torsion in Lyra geometry provides interesting isotropic and anisotropic cosmological solutions \citep{lyracosmology,lyracosmology1,lyracosmology2}, as well as modified Kaluza-Klein models \cite{lyrakk1,lyrakk2}.

We construct isotropic and anisotropic branes in $5D$ as vacuum solutions of tensor-vector dynamics. We show that the Lyra background field can be a source for the bulk cosmological constant. By assuming the background vector as the derivative of a scalar field and an independent bulk cosmological constant, pure geometric thick brane solutions are found. 

The torsion effects on the bulk bosonic and fermionic fields are also investigated. The presence of torsion enables non-minimal couplings providing dissipative and massive terms. Unlike in usual Einstein-Cartan theories, where the torsion breaks the $U(1)$ gauge symmetry, by using the associated symmetric connection, the gauge symmetry is preserved and a scalar coupling similar to a derivative dilaton-Maxwell term is allowed. For massless fermion, besides the modification in the spinor connection, torsion allows a geometric Yukawa-like interaction relating torsion, metric and the fermion. These derivative terms strong modifies the Kaluza-Klein (KK) modes nearby the brane.

This work is organized as follows: in section (\ref{section2})  we review the main properties of torsion in Lyra geometry, discussing the gravitational action and the deriving the modified Einstein equation. In section (\ref{section3}) thin and thick braneworld solutions are obtained with a background torsion and their geometric features {\color{red} are} studied. In section (\ref{section4}), the bulk fields dynamics is analyzed in both isotropic and anisotropic torsion solutions. Finally, section (\ref{section6}) is devoted to additional comments and possible future investigations.

Throughout this text we adopt the mostly minus metric signature $(+,-,\cdots,-)$. Further, we use capital latin for the bulk space-time indexes $A= 1,\cdots , D$, and greek indexes for the brane coordinates $\mu= 0, \cdots , 3$.

\section{Torsion in Lyra geometry}
\label{section2}

In this section we review the definitions and the main properties of the torsional Lyra geometry.
Moreover, the modified Einstein equation is obtained.

Firstly, let us introduce an affine connection in spacetime. In a parallel transport from the point $x^{A}$ to $x^{A}+dx^{A}$ , a vector $v^{M}$ changes as $dv^{M}=-\Gamma_{AB}^{M}v^{A}dx^{B}$. The connection coefficients induce a covariant derivative in the form $\nabla_{A}v^{M}=\partial_{A}v^{M}+ \Gamma_{AB}^{M}v^{B}$. The antisymmetric part of the connection coefficients yields the torsion tensor  $T^{M}_{AB}:=\Gamma_{AB}^{M}-\Gamma_{BA}^{M}$. Given a metric $g_{NP}$, the connection leads to the so-called non-metricity tensor $Q_{MNP}$ defined as $Q_{MNP}=\nabla_M g_{NM}$. Generically, an affine connection can be written as \cite{affine,affine2}
\begin{eqnarray}
\Gamma^{M}_{AB}=\lbrace^{M}\ _{AB}\rbrace + K^{M}\ _{AB} + L^{M}\ _{AB}.
\end{eqnarray}
The first term is the Levi-Civita connection given by  $\lbrace^{M}\ _{AB}\rbrace:=\frac{g^{MN}}{2}(\partial_A g_{NB}+ \partial_B g_{NA}- \partial_N g_{AB})$ \cite{hehl}, which is a torsion free connection $T^{M}_{AB}=0$ satisfying the metric compatibility condition $\nabla_{A}g_{BC}=0$. The second term is the contorsion tensor $K^{M}\ _{AB}= \frac{1}{2}T^{M}\ _{AB} + T_{(A}\ ^{M}\ _{B)}$ and the third term is the disformation tensor $L^{M}\ _{AB}=\frac{1}{2}g^{MC}(-Q_{ACB}-Q_{BCA}+Q_{CAB})$ \cite{affine, affine2}. In Weyl geometry, the connection considered is torsion-free $(T^{M}\ _{AB}=0)$ and the nonmetricity tensor is proportional to a geometric vector $a_M$ by $Q_{MNP}=a_M g_{NP}$ \cite{weyl,weyl1,weyl2,weyl3}. Thus, the disformation tensor in the Weyl geometry has the form $L^{M}\ _{AB}=\frac{1}{2}(\delta^{M}_A b_B + \delta^{M}_B b_A - g_{AB}b^{M})$.

A special example of torsion space-time is given in the so-called Lyra geometry, where the torsion tensor $T^{M}\ _{AB}$  is generated by a vector field $b_1=b_M dx^M$ in the form \cite{Sen-1971}
\begin{equation}
T^{M}_{AB}=\widetilde{\Gamma}_{AB}^{M}-\widetilde{\Gamma}_{BA}^{M}=-\frac{1}{2}\left(\delta^{M}_A b_B - \delta^{M}_B b_A\right),
\end{equation}
whereas a vector $V^{M}$ is modified by a parallel transport as 
\begin{equation}
\delta V^M = -\Omega(x)\widetilde{\Gamma}_{AB}^{M}V^Adx^B.
\end{equation}
The possible torsion geometries can be classified by means of the components of the torsion tensor.
A general torsion tensor can be decomposed as \cite{shapiro}
\begin{equation}
T_{MNP}=\frac{1}{(D-1)}(g_{MP}T_N - g_{MN}T_P)-\frac{1}{(D-3)2!}\epsilon_{MNPQ_{1}\cdots Q_{D-3}}S^{Q_{1}\cdots Q_{D-3}} + q_{MNP},
\end{equation}
where $T_{A}:=T^{B}_{AB}$ is the trace vector, $S^{Q_{1}\cdots Q_{D-3}}:=\epsilon^{MNPQ_{1}\cdots Q_{D-3}}T_{MNP}$ is the torsion axial tensor, $q_{MNP}$ is the null trace tensor, and $D$ is the space-time dimension \citep{hellayel,shapiro}. 
Therefore, the Lyra torsion is a pure trace vector torsion geometry with $T_M =\frac{(D-1)}{2}b_M$, where the torsion vector $b_M$ is called the displacement vector \cite{Sen-1971}.

Besides torsion, Lyra geometry has a rich notion of scale invariance gauge. In fact, in addition to a coordinate system $x^{M}$, a reference frame is endowed with a scale function $\varphi$, i.e., $(\varphi, x^{M})$, such that a vector $V^{M}$ transforms as $V'^{M}=\Omega \frac{\partial x'^{M}}{\partial x^{N}}V^{N}$ and $\Omega:=\frac{\varphi(x')}{\varphi(x)}$ \cite{Sen-1972}. The scale function $\Omega$ is an extra geometric gauge freedom which induces a conformal transformation in the form   
$ds^{2} =\Omega^{2}(x) g_{MN}dx^{M} dx^{N}$ \cite{Sen-1971}. The tensor vector $b_M$ changes under a scale transformation as $b'_M=\Omega^{-1}(b_M + \varphi^{-1}\partial_M(\log \Omega^2))$ \cite{Sen-1972}.

Assuming the Lyra connection is compactible with the metric $\tilde{\nabla}_{A}g_{MN}=0$, we can obtain an associated torsion-free Lyra connection $\nabla_{A}^{L}$ whose components are given by \cite{Sen-1971},
\begin{equation}
\label{Lyraconnection}
\Gamma_{AM}^M = \Omega^{-1}\lbrace_{AM}^M\rbrace + \frac{1}{2}(\delta_A^M b_B + \delta_B^M b_A - g_{AB}b^M).
\end{equation}
Note that for the gauge choice $\Omega=1$ the torsion-free Lyra connection is similar to the Weyl connection \cite{Sen-1972}.

The Lyra connection modifies the Ricci tensor in the form $R_{MN}^L=R_{(MN)}^L+R_{[MN]}^L$, where the symmetric part is given by \cite{Sen-1972}
\begin{equation}
    R_{(MN)}^L=\Omega^{-2}R_{MN}+\frac{\Omega^{-1}}{2}(2\nabla_{(M}b_{N)}+\nabla_{C}b^c g_{MN})-\frac{1}{2}(b_M b_N -b_C b^C g_{MN}),
\end{equation}
where $R_{MN}$ is the Ricci tensor obtained from the Levi-Civita connection and the skew-symmetric part has the form \cite{Sen-1972} 
\begin{equation}
    R_{[MN]}^L=\Omega^{-1}f_{MN},
\end{equation}
where $f_{MN}:=\partial_M b_N - \partial_N b_M$ \cite{Sen-1972}. The appearance of a skew-symmetric term in the Ricci tensor is an ubiquitous feature of torsion theories and in Einstein-Cartan models is related to the spin interaction \cite{hehl}. In Lyra torsion model the $R_{[MN]}^L$ term seems linked to the dynamics of the torsion vector $b_M$.

The Ricci scalar curvature derived from the Lyra connection $R^L :=g^{MN}R_{MN}$ for $\Omega=1$ can be written as \cite{Sen-1971},
\begin{equation}\label{k-gauge1}
R^L = R + \frac{1}{4}(D - 2)(D - 1)b^{A}b_{A}   + (D - 1)\nabla^{A}b_{A}
\end{equation} 
where $D$ is space-time dimension, $R$ is the Ricci scalar build from the torsion-free Levi-Civita connection and the capital letters represent the indexes in bulk. Likewise Weyl geometry, the torsion tensor can be regarded as an external field modifying the space-time curvature. Notwithstanding, note that the skew-symmetric component of the Lyra-Ricci tensor vanishes when contracted with the metric.

The presence of space-time torsion enlarges the possible gravitational dynamics, by allowing new interactions and possibly considering both the metric and connection as independent variables \cite{hehl}. Nonetheless, in the present work we consider the torsion vector as a background field and assume a metric theory described by an extension of the Einstein-Hilbert gravitation action in the form
\begin{equation}
S_g =\frac{1}{\kappa_g}\int \Omega^D R^{L}\sqrt{|g|}dx^1\cdot \cdot \cdot dx^D, 
\end{equation}
which yields the modified Einstein equation
\begin{equation}\label{einsteinmodified}
R_{MN} - \frac{1}{2}g_{MN}R = -\frac{\Omega^2}{4}(D-2)(D-1)(b_M b_N - \frac{1}{2} g_{MN} b_C b^C).
\end{equation}
The Einstein equation (\ref{einsteinmodified}) may be interpreted as a tensor-vector-scalar theory of gravity or only a tensor-scalar theory by assuming $b_M=\partial_M h$. This theory is similar to the Brans-Dicke gravity in some special regimes \cite{Sen-1971}.

It is worthwhile to mention that a kinetic term for the torsion vector can be obtained as
\begin{equation}
    S_b = \alpha\int{\Omega^D R_{[MN]} R^{[MN]}\sqrt{|g|}dx^1\cdot \cdot \cdot dx^D},
\end{equation}
which yields a Lagrangian similar to the dilaton-Maxwell term $\alpha \Omega^{D-2}f_{MN}f^{MN}$. Accordingly, considering only linear terms for the curvatures, i.e., without high-order terms in the gravitational Lagrangian, it yields background torsion vector. Further, for $b_M=\partial_M h$, i.e., in the tensor-scalar regime, the term $R_{[MN]} R^{[MN]}$ vanishes identically.

The effects of the torsion in Lyra geometry were investigated in black holes modified solutions \cite{Sen-1971}, cosmology \cite{lyracosmology1,lyracosmology2} and Kaluza-Klein (KK) models \cite{lyrakk1,lyrakk2}. In next section we explore the torsion effects in warped braneworld models assuming the torsion as a source for the thick brane configurations. 


\section{Warped braneworld solutions}
\label{section3}

In this section we study some brane geometries which are solutions of the modified Einstein equation (\ref{einsteinmodified}).

For a covariant constant displacement vector $\nabla_M b^{M}=0$ the torsion provides the bulk cosmological constant. In fact, 
the norm $b_{M}b^{M}$ is a global constant throughout the bulk and hence we can define the bulk cosmological constant $\Lambda$ as
\begin{equation}
\Lambda = \frac{(D-1)(D-2)}{8}b_{M}b^{M}.
\end{equation}
Thus, for an $AdS_D$ bulk, $b_M$ ought to be a spacelike vector.

Consider a warped brane metric in the form
\begin{equation}
ds_5^2 = e^{-2A(y)}\eta_{\mu\nu}dx^{\mu}dx^{\nu} - dy^2.
\end{equation}  
For $\Lambda= - 3b^2 /2 < 0$ let us consider $b_M = (0, 0, 0, 0, b_4)$. The vacuum Einstein equation in the extra dimension yields 
\begin{equation} \label{thinbrane}
A(y)= \pm\left(\frac{b_4}{4}\left| y\right|\right) + A_0,
\end{equation}
providing a thin 3-brane immersed in the $AdS_5$ RS model \cite{rs1}. The norm of $y$ in (\ref{thinbrane}) was chosen to preserve $Z_2$ symmetry. Since $b^2$ is a source for a negative bulk cosmological constant, this configuration leads to the usual RS model, wherein a thin brane is embedded into a $AdS_5/Z_2$ bulk. The relationship between the brane tensions and the bulk cosmological constant is the same as in the RS models. Accordingly, the Lyra torsion vector only provides an origin for the bulk cosmological constant of the RS models.

For a spacelike vector $b_A=(0,b_1,b_2,b_3,b_4)$, the contravariant components are $b^{i}=e^{2A}b_i$. Therefore, the background vector $b^{i}$ is exponentially suppressed on the 3-brane at the origin. Since the background vector can couples with the standard model fields, as we will discuss in next section, the suppression of this vector might explain the small departure of Lorentz symmetry on the brane \cite{kostelecky1}. 

An interesting interpretation of the torsion in Lyra spacetimes is achieved for $b_M = \partial_M h(x)$ from which the Einstein equation with a bulk cosmological constant reads
\begin{equation}
\label{modifiedeinsteinscalar}
G_{MN}  = - 3\tilde{T}_{MN} + \Lambda g_{MN},
\end{equation} 
where $\tilde{T}_{MN}=\left(\nabla_M h \nabla_M h - \frac{1}{2} g_{MN}\nabla^Ch\nabla_Ch\right)$ is analogous to the stress energy tensor of a massless real scalar field $h(x)$. Assuming that the Lyra scalar field $h(x)$ does not propagate on the brane, the free field equation of motion $g^{M N}\nabla_M\nabla_N h = 0$ yields $-h''(y)+4A'(y)h'(y)=0$
whose solution is $h = \int e^{(4A + c_1)}dy + c_2$.

The Einstein equation in the extra dimension yields
\begin{equation}
\label{warpfactorequation}
6A'^2 = -\frac{3}{2}h'^2 - \Lambda,
\end{equation}
thus, the torsion provides a negative term in Eq.(\ref{warpfactorequation}), which possesses solution provided $h'\leq (2\Lambda/3)^{1/2}$.

For the free Lyra scalar $A =\ln{\Big[\frac{-2\Lambda}{h'_0}\sech^2\left(2\sqrt{\frac{-2\Lambda}{3}}y\right)\Big]^\frac{1}{8}}$,  the Lyra scalar field has the form $h=tg^{-1}\left(\tanh\Big[2\sqrt{\frac{-2\Lambda}{3}}y+c_2\Big]\right).$ Therefore, a free Lyra scalar field exhibits a kink-like profile with a localized energy density around the brane at $y=0$ and a nonvanishing value asymptotically. In a similar form, accelerated cosmological solutions in 4D were found in bounce models generated by a Weyl scalar \cite{Novello}.  This free field configuration provides a divergent warp factor and thus, an additional mechanism is required in order to trap gravity. Note, however, that a constant derivative $h'=h_0$ satisfies the free field equation with a constant energy density and negative pressure. Thus, the Lyra scalar provides a positive cosmological constant term which opposes to the bulk cosmological constant. 

\begin{figure} [htb]
       \begin{minipage}[b]{0.48 \linewidth}
           \includegraphics[width=\linewidth]{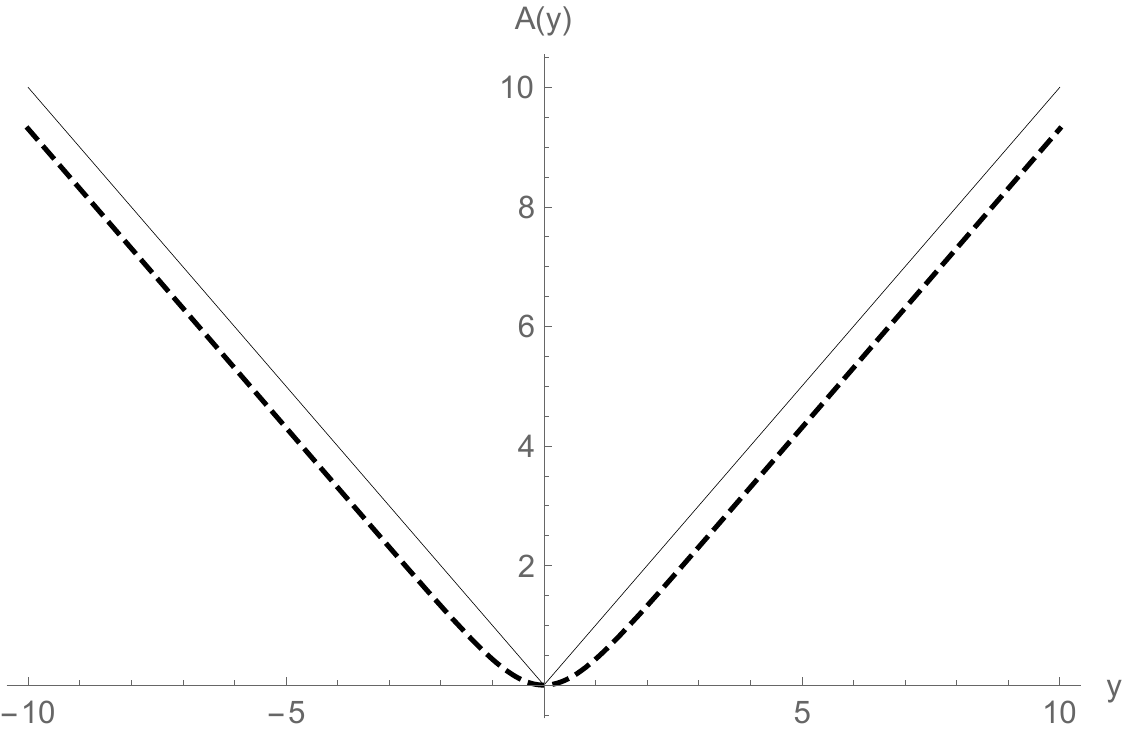}\\
           \caption{Warp function for $\lambda=1$ (dashed line) and  for the RS model (thin line).}
          \label{warpfactor1}
       \end{minipage}\hfill
       \begin{minipage}[b]{0.48 \linewidth}
           \includegraphics[width=\linewidth]{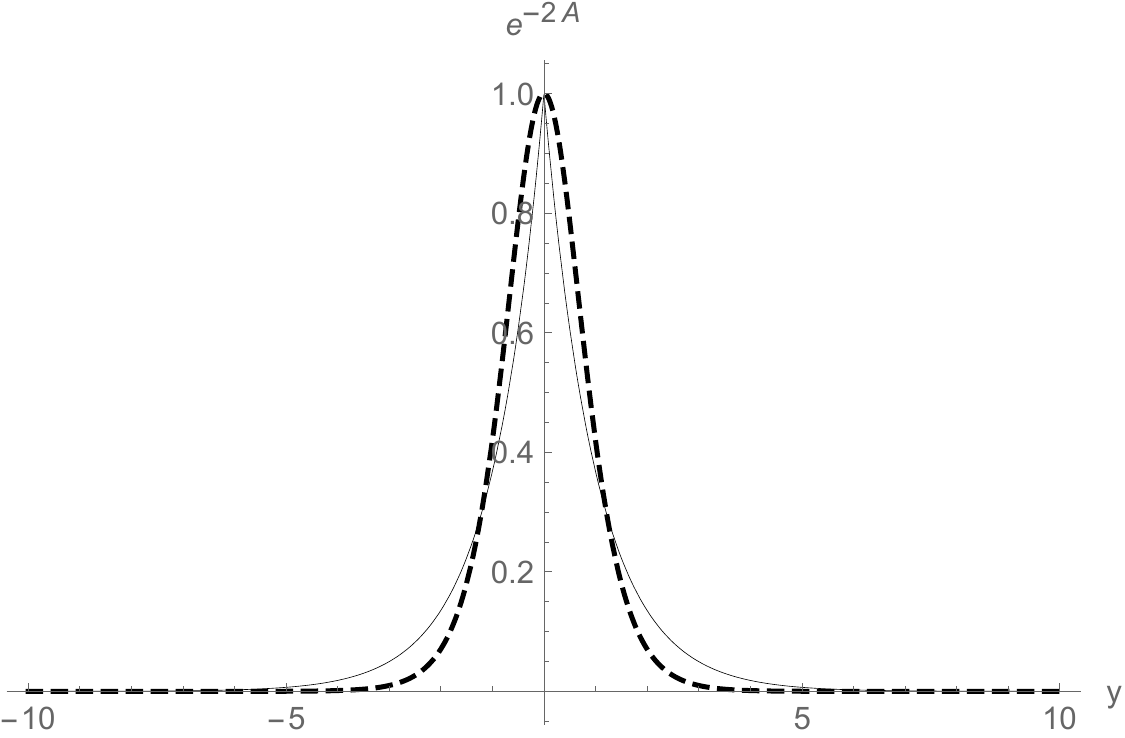}\\
           \caption{Warp factor for $\lambda=1$ (dashed line) and  for the RS model (thin line).}
           \label{warpfactor2}
       \end{minipage}
\end{figure}

The profile of the free Lyra scalar motivated us to seek for a background Lyra scalar localized near the origin in the form
\begin{equation}
\label{lyrascalarlocalized}
h(y)=\frac{1}{\lambda}\sqrt{\frac{-2\Lambda}{3}}tg^{-1}\left(\tanh\lambda y\right).
\end{equation}
Using this \textit{ansatz}, the Einstein equations (\ref{warpfactorequation}) provides the warp factor
\begin{equation}
\label{isotropicwarpfactor2}
A(y)=\frac{1}{\lambda}\sqrt{\frac{-\Lambda}{6}}\log \cosh \lambda y,
\end{equation}
where $\lambda$ measures the width of the Lyra scalar non-constant region. In Figure (\ref{warpfactor1}) we compare the RS $Z_2$-orbifolded solution $A(y)=k|y|$ (thin line) to the smooth warp factor in Eq.(\ref{isotropicwarpfactor2}). Thus, remarkably, this pure-geometric braneworld scenarios is equivalent to the thick brane models.

\begin{figure} 
       \begin{minipage}[b]{0.48 \linewidth}
           \includegraphics[width=\linewidth]{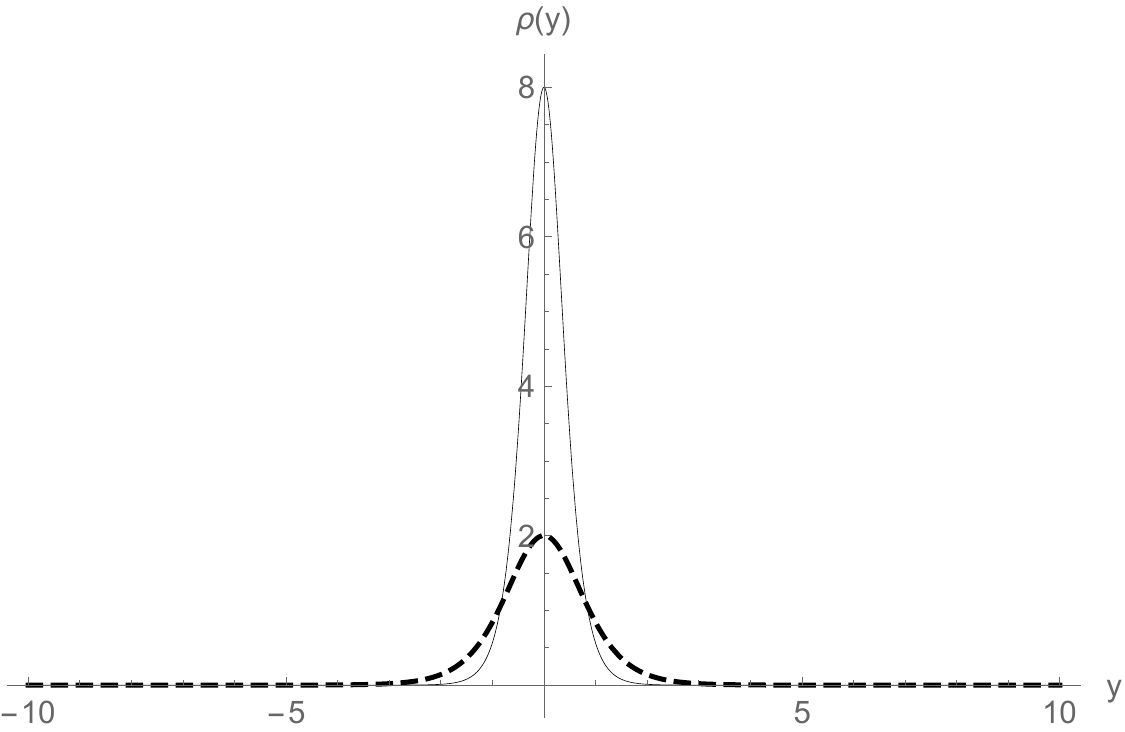}\\
           \caption{Energy density for $\lambda=1$ (dashed line) and  $\lambda=2$ (thin line) using $\Lambda=-\lambda^2$.}
          \label{energydensity}
       \end{minipage}\hfill
       \begin{minipage}[b]{0.48 \linewidth}
           \includegraphics[width=\linewidth]{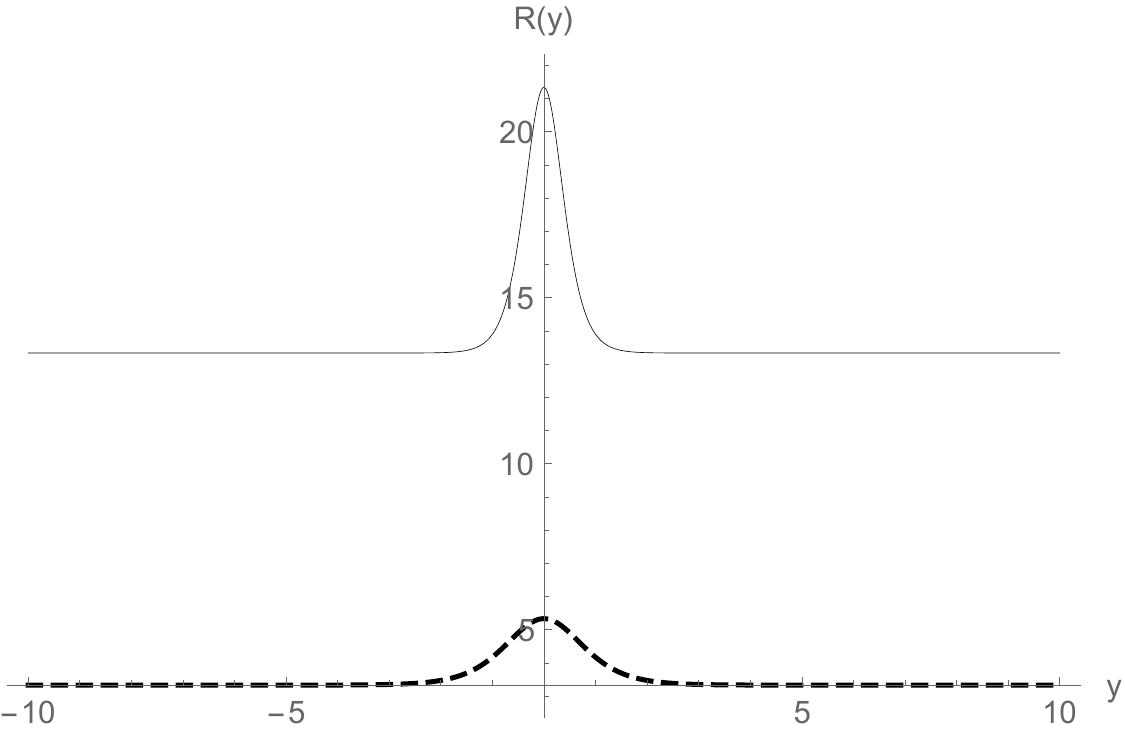}\\
           \caption{Ricci scalar for $\lambda=1$ (dashed line) and $\lambda=2$ (thin line) using $\Lambda=-\lambda^2$.}
           \label{curvature}
       \end{minipage}
   \end{figure}

The resulting geometry is a family of asymptotically $AdS_5$ spacetimes with warped metrics parametrized by the ratio between $\Lambda$ and $\lambda$ in the form 
\begin{equation}
ds^2=\sech^{2\sqrt{-\Lambda/6\lambda^2}}(\lambda y)\eta_{\mu\nu}dx^\mu dx^\nu - dy^2.
\end{equation}
This braneworld solution, including the scalar field pattern, is rather similar to one found in teleparallel $f(T)$ gravity \citep{Yang, allan1, allan2}.  The profile of the energy density and the scalar curvature are shown, respectively in Figures (\ref{energydensity}) and (\ref{curvature}). The torsion term in Einstein equation provides a negative pressure that smooths the brane.

\section{Bulk fields} \label{section4}

In this section we investigate the properties of bulk fields in this torsion geometry. The background Lyra vector provides non-minimal couplings whose effects are analyzed.

\subsection{Scalar field}

In order to study the behavior of scalar field consider the action for a minimally coupled scalar field
\begin{equation} \label{scalar-action}
S_{\Phi}=\frac{1}{2}\int d^5x \sqrt{-g}\left(g^{MN} \nabla_{M}^{L} \Phi \nabla_{N}^L \Phi \right).
\end{equation}
Since $\nabla^{L}_M \Phi = \partial_M \Phi$, then a minimally coupled scalar field does not probe the torsion geometry. To do so, we suppose non-minimall couplings of the form
\begin{equation}
\label{scalarfieldeom}
\frac{1}{\sqrt{-g}}\partial_M \left( \sqrt{-g} g^{MN}  \partial_N\Phi \right)+\frac{3\alpha}{4}T^{P}\partial_P \Phi-\zeta R\Phi = 0.
\end{equation}
The torsion coupling $\frac{3}{4}T^{P}\partial_P \Phi$ provides a dissipative-like term depending on the background vector $b_M$ whose  coupling constant is $\alpha$. The non-minimal coupling to the Ricci scalar leads to indirect effects of the torsion on the scalar field.

Consider the isotropic brane determined by the warp factor in Eq.(\ref{isotropicwarpfactor2}). The Kaluza-Klein decomposition
$\Phi(x,y)=\tilde{\Phi}(x)\chi(y)$, where the scalar field on the brane satisfies $\partial_\mu \partial^{\mu}\tilde{\Phi}=m^2 \tilde{\Phi}$, yields an EOM from Eq.(\ref{scalarfieldeom}), namely
\begin{equation}
\label{scalarisotropiceom}
\chi''- \left(4\sqrt{-\Lambda}\tanh\lambda y-\frac{3\alpha}{2}b_y \right)\chi' +\Big\{ m^2 \cosh^{p}\lambda y-2\zeta\Lambda \left(\frac{5}{3}+{\sech^{2}\lambda y}\right)\Big\}\chi=0,
\end{equation}
where $p:=\sqrt{-\Lambda}/\lambda$ is a dimensionless parameter.
Near the brane, the massive modes satisfy $\chi''(y)+\frac{3}{2}\lambda\sqrt{\frac{2}{\kappa_5}} \chi'(y) + m'^2 \chi(y)=0$, where $m'^2 = m^2 -\frac{16\zeta\Lambda}{3}$. Thus, the torsion coupling adds a damping factor $\exp(-\frac{3}{4}\lambda\sqrt{\frac{2}{\kappa_5}}y)$ into the massive modes around the brane.

For $\Lambda=-6\lambda^2$, the change of coordinate $z=(1/\lambda)\sinh \lambda y$ turns the metric into the conformal form
$ds^2=\frac{1}{1+\lambda^2 z^2}(\eta_{\mu\nu}dx^\mu dx^\nu - dz^2)$. In these coordinates, the KK equation (\ref{scalarisotropiceom}) takes a Schr\"{o}dinger-like form $-\ddot{\Psi}+{U_{\phi}(z)}\Psi=m^2 \Psi$, where 
\begin{eqnarray}
\label{scalarpotentialequation}
U_\Phi &=& \frac{3\lambda^3 z^2}{(1+\lambda^2 z^2)^2} -\frac{3\lambda}{2(1+\lambda^2 z^2)} - \frac{3\alpha \lambda^3 (z+4\lambda^2 z^3))}{(1+\lambda^2 z^2 + \lambda^4 z^4)}\nonumber\\
&+&\frac{9\lambda^2}{4}\left(\frac{z}{1+\lambda^2 z^2}-\frac{\alpha}{1+\lambda^2 z^2 + 2\lambda^4 z^4}\right) + \frac{2\lambda^2 \zeta}{1+\lambda^2 z^2}\left(\frac{5}{3}+ \frac{1}{1+\lambda^2 z^2}\right),
\end{eqnarray}
the dot stands for $d/dz$ and $\chi=\sqrt{1+\lambda^2 z^2}e^{-\frac{3\alpha}{4}h(z)}\Psi(z)$. The effective of the dissipative coupling is controlled by $\alpha$ whereas the strength of the non-minimal coupling is controlled by $\zeta$. Note that the dissipative coupling leads to an odd term in the third term of the potential (\ref{scalarpotentialequation}). Thus, the dissipative coupling breaks the
$Z_2$ symmetry on the potential whereas the non-minimal coupling preserves such symmetry.

\begin{figure}[htb] 
       \begin{minipage}[b]{0.48 \linewidth}
           \includegraphics[width=\linewidth]{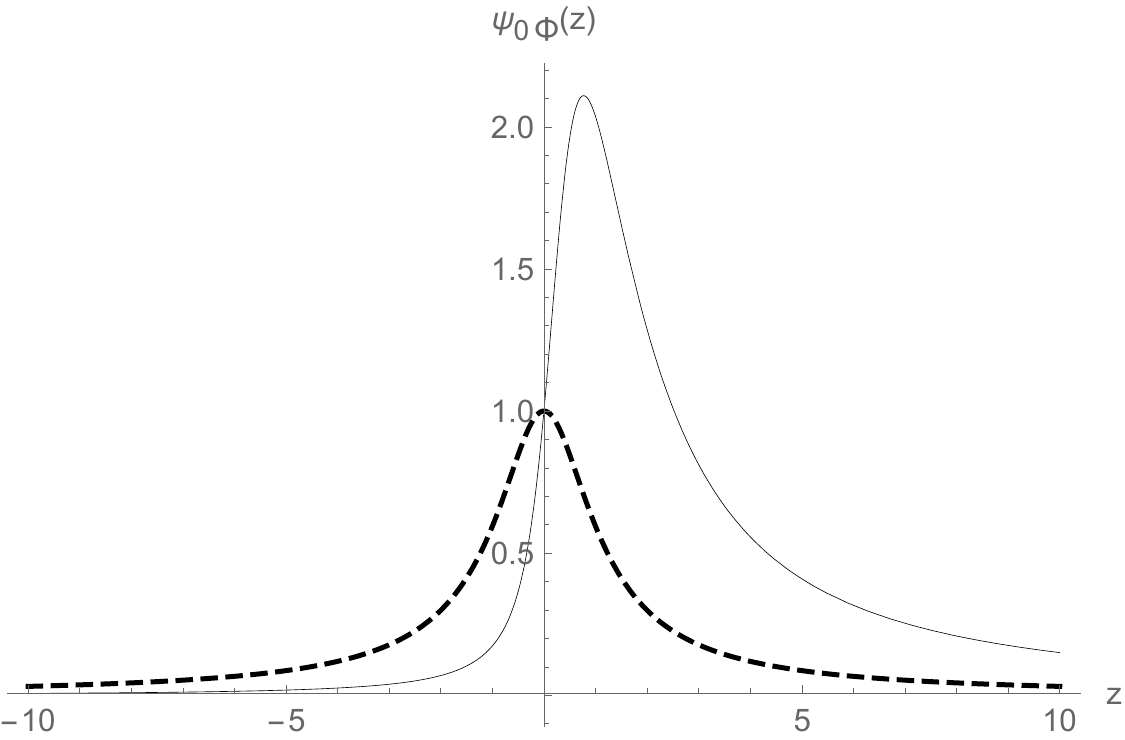}\\
           \caption{Scalar field massless mode for $\lambda=1$ with the torsion coupling $\alpha=1$ (thin line) and without the torsion coupling $\alpha=0$ (dashed line).}
          \label{scalarzeromode}
       \end{minipage}\hfill
       \begin{minipage}[b]{0.48 \linewidth}
           \includegraphics[width=\linewidth]{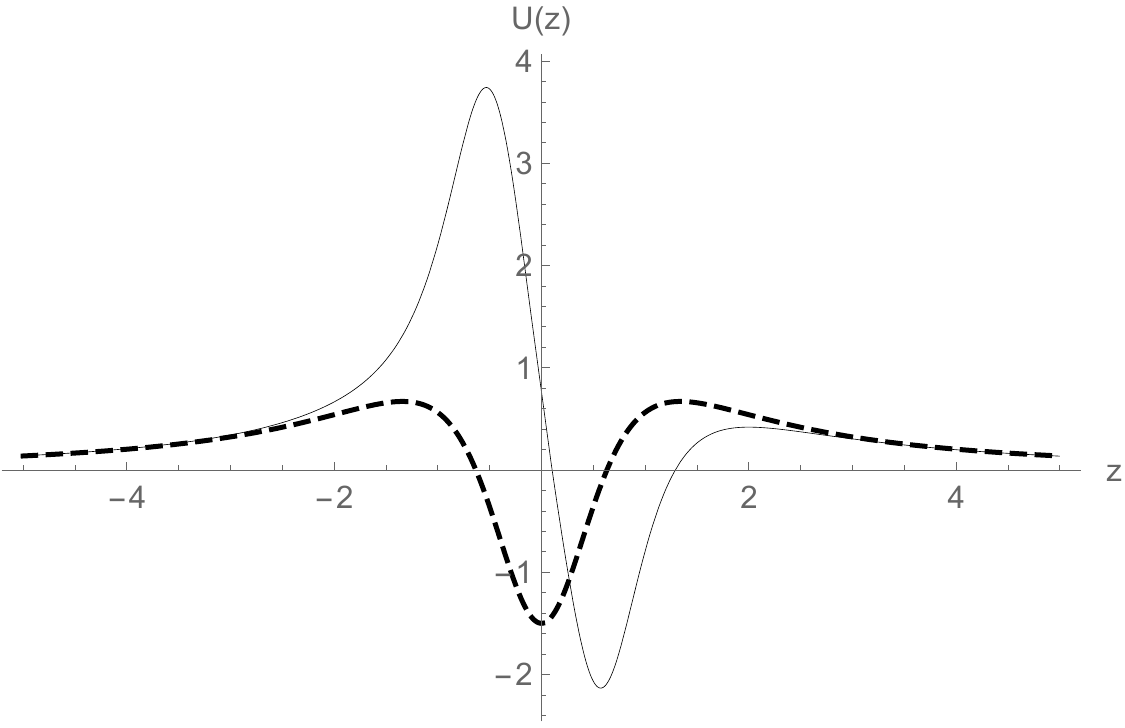}\\
           \caption{Potential for the scalar field with dissipative torsion coupling $\alpha=1$ (thin line) and without $\alpha=0$ (dashed line). The background  Lyra vector breaks the $Z_2$ symmetry.}
           \label{scalarpotentialdissipative}
       \end{minipage}
   \end{figure}

For $\zeta=0$, the analogue Schr\"{o}dinger potential $U_\Phi$ has a SUSY-like structure $U_\Phi = W^2 -\dot{W}$, where the superpotential is $W=\frac{3}{2}\left(\dot{A}-\frac{1}{2}\dot{h}\right)$. In this regime, this potential enables a normalized massless mode
\begin{equation}
\Psi_0 = \frac{N}{(1+\lambda^2 z^2)^{3/4}}e^{(3\alpha/2)\tan^{-1}\left(\frac{\lambda z}{\sqrt{1+\lambda^2 z^2}}\right)}.
\end{equation}
The scalar massless mode for $\zeta=0$ is sketched in fig.(\ref{scalarzeromode}). Note that the dissipative coupling shifts the massless mode from the origin. As shown in the fig.(\ref{scalarpotentialdissipative}), the shift in the massless mode is due to the break of the $Z_2$ symmetry on the potential.


\begin{figure}[htb] 
       \begin{minipage}[b]{0.48 \linewidth}
           \includegraphics[width=\linewidth]{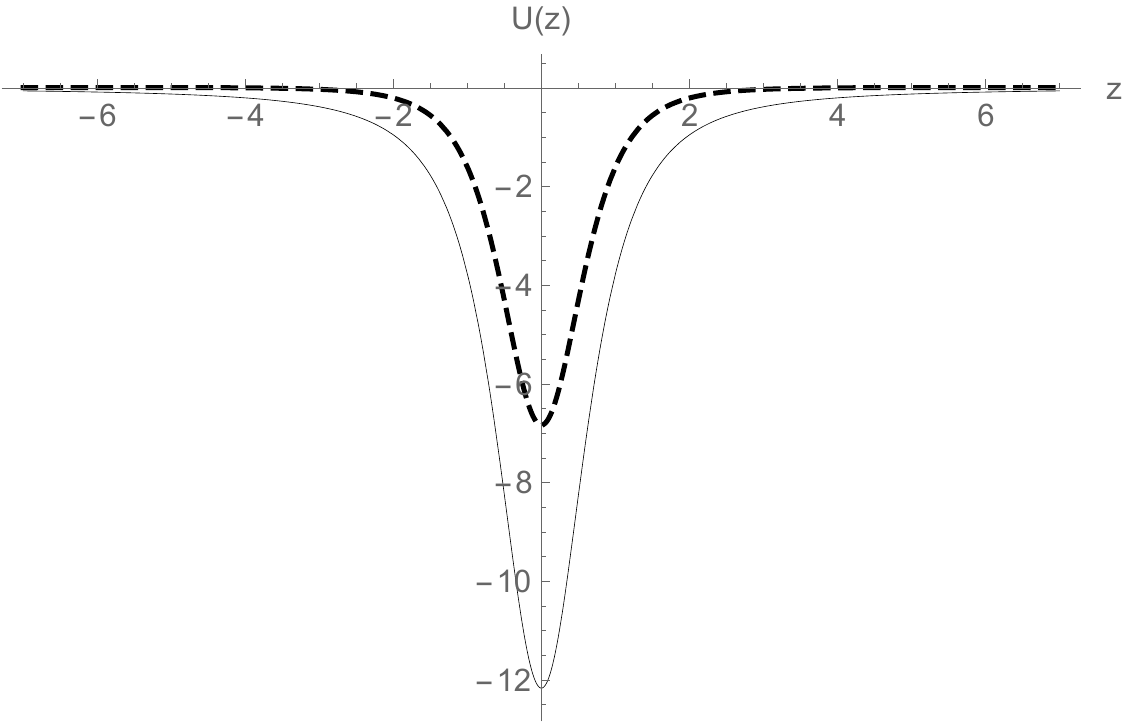}\\
           \caption{Potential for $\lambda=1$ with the nonminimall coupling $\zeta=-1$ (thin line) and without $\zeta=0$ (dashed line).}
          \label{scalarpotentialnonminimal2}
       \end{minipage}\hfill
       \begin{minipage}[b]{0.48 \linewidth}
           \includegraphics[width=\linewidth]{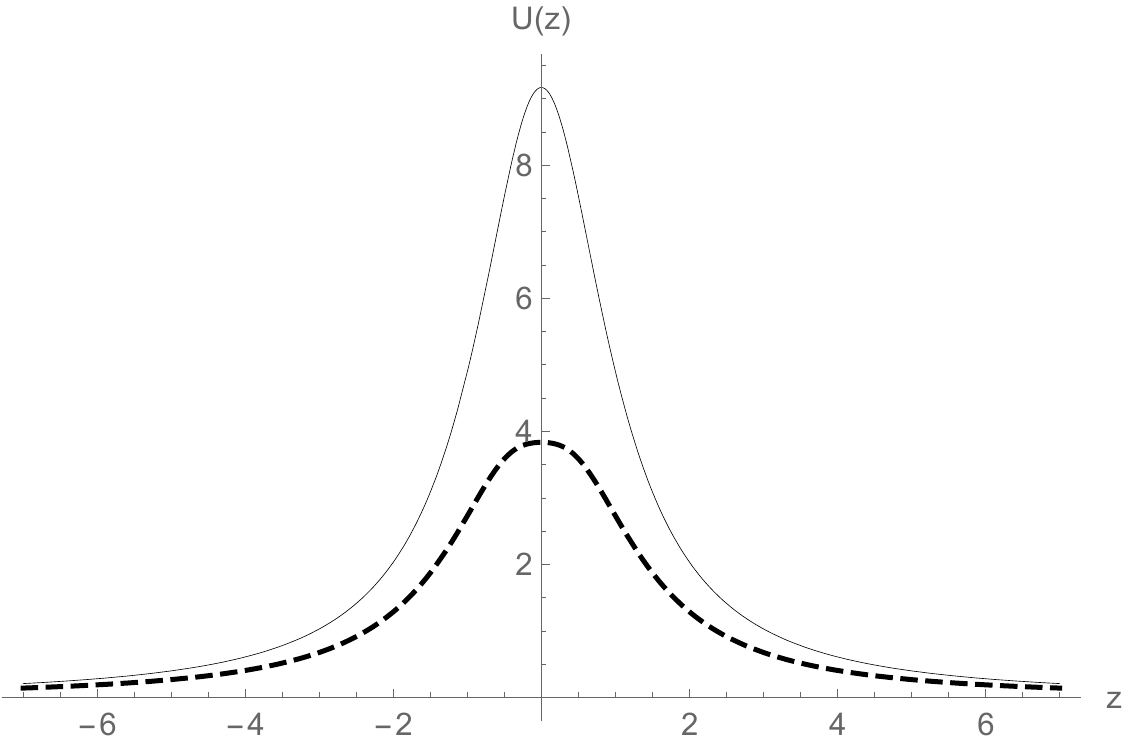}\\
           \caption{Potential for $\lambda=1$ with the nonminimall coupling $\zeta=1$ (thin line) and without $\zeta=0$ (dashed line).}
           \label{scalarpotentialnonminimal1}
       \end{minipage}
   \end{figure}

For $\zeta\neq 0$ and $\alpha=0$, albeit the massive tower remains not localized. 
Further, for a scalar field without the torsion coupling, the potential exhibits a symmetric profile. For $\zeta<0$, the nonminimal coupling leads to a potential well around the origin, as shown in Fig.(\ref{scalarpotentialnonminimal2}). On the other hand, for $\zeta>0$, the potential exhibits a symmetric barrier at the origin, as depicted in the Fig.(\ref{scalarpotentialnonminimal1}).


\subsection{Gauge field}
\label{gaugefield}

We begin considering a general gauge invariant Lagrangian for a vector field in the form
\begin{equation}
\mathcal{L}_A = \mathcal{L}_{A,T}+\mathcal{L}_{A,S}
\end{equation}
where the trace-torsion vector coupling is
\begin{equation}
\label{nonminimalgaugecoupling}
\mathcal{L}_{A,T}:=-\frac{1}{4}\sqrt{g}e^{\xi(T_C T^C)}g^{MN}g^{PR}F_{MP}F_{NR}
\end{equation}
and the axial coupling is
\begin{equation}
\mathcal{L}_{A,T}:=\xi_1 \sqrt{g}S^{MP}S^{NR}F_{MP}F_{NR}.
\end{equation}
The field strength is defined as $F_{MP} = \nabla_M^{L} A_P - \nabla_P^{L} A_M$ and $\nabla_P^{L}$ is the torsion-free connection. The presence of torsion usually breaks the gauge symmetry \cite{Casanagauge,Casanalyragauge}. Nevertheless, the existence of an associated torsion-free connection in Lyra geometry preserves the $U(1)$ gauge symmetry. Further, in Lyra geometry, the tensor $S_{MN}$ vanishes and then we consider the $\mathcal{L}_{A,T}$ term only.  The coupling between the gauge field and the background vector $b_C$ provides a birefringence interaction for the vector field. The norm of the trace torsion vector $T_C$ can be interpreted as a kind of dilaton interacting with the spin one field. For $b_C =\partial_C h$, this term can be regarded as a derivative dilaton-Maxwell theory. Note that for $\xi=0$, the usual Maxwell lagrangian is recovered. The nonminimal coupling constant $\xi$ has dimension of $L^2$. The resulting equation of motion is 
\begin{equation} \label{gauge-move-eq}
\frac{1}{\sqrt{g}}\partial_M\left(e^{-4\xi b^2}\sqrt{g}g^{MN}g^{PR}F_{NR} \right) = 0,
\end{equation}
where $b_M b^M =-b^2$, for $b_M = (0,\Vec{0},b)$.

In the isotropic scenario with the warp factor given by Eq.(\ref{isotropicwarpfactor2}), performing the KK dimensional reduction $A_\mu (x,y)=\tilde{A}_\mu (x) \Phi(y)$ and choosing $\partial_\mu A^\mu=A_y=0$, we obtain the KK equation
\begin{equation}
\label{gaugeisotropickkequation}
\Phi''+(-4A' - 4\xi \dot{(b^2)})\Phi' + e^{2(A-2\xi b^2)}\Phi=0
\end{equation}
where the KK mass arises from $\partial_\mu \partial^\mu A^{\nu}=m^2 A^{\nu}$ and $b^2=b_C b^C$. The massless mode is given by
\begin{equation}
\Phi_0 = \int^{y}e^{4(A+\xi b^2)}dy' + c.
\end{equation}
Although the first term diverges, a constant massless mode $\Phi_0=c$ is localized provided that $\int{e^{-\xi b^2} dy}$ converges, as for $b^2 = h'^2$ in Eq.(\ref{lyrascalarlocalized}).

Performing the change of coordinate $z=\int{e^{A-2\xi b^2}dy}$, the Eq.(\ref{gaugeisotropickkequation}) turns into $-\ddot{\Psi_{A}}+U_A (z')\Psi_A =m^2 \Psi_A$, where 
\begin{equation}
U_A = \frac{9}{4}\left(\dot{A}+2\xi \dot{b^2}\right)^2 -\frac{3}{2}\left(\ddot{A}+2\xi \ddot{b^2}\right).
\label{gaugepotential}
\end{equation}
The inclusion of the nonminimal coupling preserves the stability of the KK modes, for the potential keeps the form $U_A = W_{A}^2 -\dot{W_A}$. 
Unlike for the scalar field, the Schr\"{o}dinger equation is not written in conformal coordinate due to the presence of the non-minimal coupling. In spite of the analytical form of the function $z(y)$ can not be found in general, the regularity and positivity of $z'(y)$ guarantees that some asymptotic and qualitative features can be addressed. Notably, since $\dot{A}=e^{-(A-2\xi b^2)}A'$ and $A'\rightarrow c$ as $z\rightarrow \pm \infty$, the potential vanishes asymptotically provided that $e^{-(A-2\xi b^2)}\rightarrow 0$ as $z\rightarrow \infty$. This condition is fulfilled for a covariant constant background Lyra vector $b_M=e^{-A}(0,0,0,0,b_4)$ and for the scalar Lyra $b_M= \partial_M h$ as well.

The corresponding massless bound state has the form
\begin{equation}
\Psi_{0A}=N_A e^{-\frac{3}{2}(A+2\xi b^2)}, 
\end{equation}
which for a covariant constant background vector provides a constant that can be absorbed into the normalization constant. For $b_M= \partial_M h$ the massless is $\Psi_{0A}=N_A e^{-\frac{3}{2}A}e^{-3\xi \left(\dot{h^2}e^{2(A-2\xi b^2)}\right)}$ which vanishes asymptotically, for $h'=e^{A}\dot{h}\rightarrow 0$ as $z\rightarrow\infty$, thereby showing a damping effect of the non-minimal torsion interaction to the massless gauge field.


\subsection{Fermion field} \label{section5}

We propose the following non-minimal coupling of the massless bulk spinor and the torsional geometry
\begin{equation} \label{fermion-action}
S_\Psi = \int {d^{5} x \sqrt{-g} \Big[\bar{\Psi} i \Gamma ^{M} D_{M} \Psi - (\epsilon_1 h^p \sqrt{R} +\epsilon_2 n^{M}\partial_M h)\bar{\Psi}\Psi\Big]},
\end{equation}
where $p$ is an integer, $n^{M}=(0,\Vec{0},1)$ is an unit vector along the extra dimension, $\Gamma^{M} = e_{\bar{M}}^{M} \gamma^{\bar{M}}$ are the Dirac matrices in a curved spacetime, $\gamma^{\bar{M}}$ are the flat Dirac matrices and the  \textit{vielbeins} $e_{\bar{M}}^{M}$ satisfy 
$g_{M N} = \eta _{\bar{M}\bar{N}}e_{M}^{\bar{M}} e_{N}^{\bar{N}}$.
The spinorial covariant derivative in torsional Lyra geometry has the form \cite{Nieh, casanaspinor,casanalyrafermion}
\begin{equation} \label{deri-covar}
D_{M} = \partial _{M} + \Omega_M^L,
\end{equation}
where $\Omega_M^L = \Omega_M + \Omega'_M$ is the torsion spinor connection, with $\Omega_M =\frac{1}{4} \Gamma_{M} ^{\bar{M} \bar{N}} \gamma _{\bar{M}} \gamma _{\bar{N}}$ being the torsion-free spinor connection and $\Gamma_{M} ^{\bar{M} \bar{N}}$ are 1-form Levi-Civita connection coefficients, i.e., 
$\omega_{\bar{N}}^{\bar{M}}=\Gamma_{M\bar{N}}^{\bar{M}}dx^{M}$. The torsion provides a new spin connection term $\Omega'_M =-\frac{1}{2}T_M$ which for $b_M=\partial_M h$ yields $\Omega'_M=-(3/2)\partial_M h$.

Further, the background vector provides a non-minimal mass term of form
\begin{equation}
F(h,g)=(\epsilon_1 h^p \sqrt{R}+\epsilon_2 h' ),
\end{equation}
which can be regarded as a kind of geometric Yukawa-like derivative interaction between the fermion, the Lyra scalar (torsion) and the metric. The coupling constants $\epsilon_{1,2}$ are dimensionless real numbers.

\begin{figure}[htb] 
       \begin{minipage}[b]{0.48 \linewidth}
           \includegraphics[width=\linewidth]{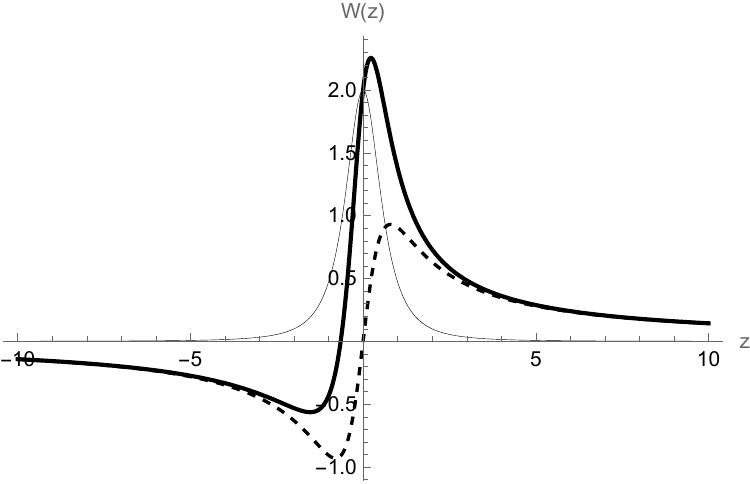}\\
           \caption{Fermion superpotential for $\lambda=1$. For $\epsilon_1=1,\epsilon_2=0$ (dashed line) the superpotential is an odd function of z, whereas for $\epsilon_1=0,\epsilon_2=1$ (thin line) and $\epsilon_1=1,\epsilon_2=1$ (thick line) the superpotential become asymmetrical.}
          \label{superpotential}
       \end{minipage}\hfill
       \begin{minipage}[b]{0.48 \linewidth}
           \includegraphics[width=\linewidth]{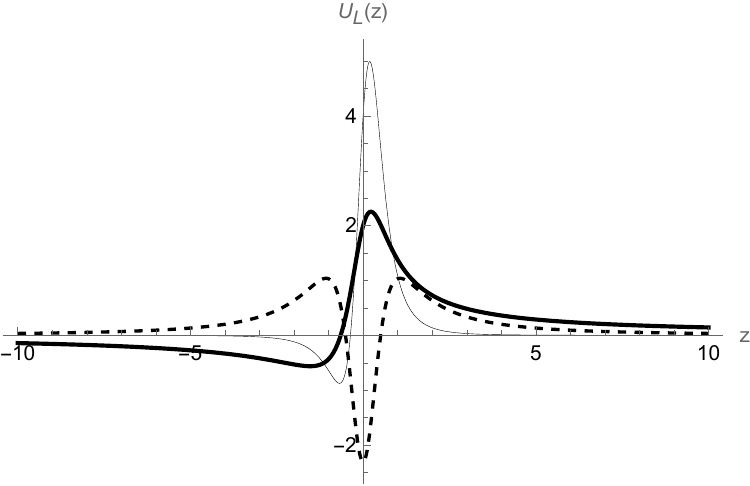}\\
           \caption{Fermion potential $U_L$ for $\lambda=1$. For $\epsilon_1=1,\epsilon_2=0$ (dashed line) the potential has the volcano shape, whereas for $\epsilon_1=0,\epsilon_2=1$ (thin line) and $\epsilon_1=1,\epsilon_2=1$ (thick line) the potential become asymmetrical.}
           \label{leftpotentialdisplay}
       \end{minipage}
\end{figure}

For the isotropic scenario in the conformal coordinate the metric can be written as $ds^2=\eta_{\bar{M}\bar{N}}\hat{\theta}^{\bar{M}}\otimes \hat{\theta}^{\bar{N}}$ where $\hat{\theta}^{\bar{M}}=e^{-A}\delta^{\bar{M}}_{M}dx^{M}$. The Cartan equation $d\hat{\theta}^{\bar{M}}+\omega^{\bar{M}}_{\bar{N}}\hat{\theta}^{\bar{N}}=0$ yields the nonvanishing 1-form connection coefficients $\Gamma^{\bar{\mu}}_{5\bar{\nu}}=-\dot{A}\delta^{\bar{\mu}}_{\bar{\nu}}$ and $\Gamma^{5}_{\bar{\mu}\bar{\nu}}=-\dot{A}\eta_{\bar{\mu}\bar{\nu}}$. Thus, the Dirac equation reads
\begin{equation}
\Big[\gamma^\mu (\partial_\mu +\frac{3}{2}b_\mu)+\gamma^z(\partial_z - 2\dot{A}-\frac{3}{2}b_z))-e^{-A}(\epsilon_1 h^p \sqrt{R}+\epsilon_2 e^{A}
\dot{h} )\Big]\Psi=0.
\end{equation}
For $\gamma^\mu (\partial_\mu +\frac{3}{2}b_\mu)\Psi=m\Psi$ and 
$\Psi(x,z)=f(z)\sum{\psi_{4R} (x)\alpha_R (z) +\psi_{4L}\alpha_L (z)}$, where $f(z)=e^{2A+\frac{3}{2}\int{b_z dz'}}$
the Dirac equation leads to
\begin{eqnarray}
\label{coupleddiracequations}
(\partial_z +  e^{-A}(\epsilon_1 h^p \sqrt{R}+\epsilon_2 e^{A}
\dot{h} ))\alpha_L &=& m \alpha_R\nonumber\\
(\partial_z - e^{-A}(\epsilon_1 h^p \sqrt{R}+\epsilon_2 e^{A}
\dot{h} ))\alpha_R &=&- m \alpha_L.
\end{eqnarray} 
By decoupling the system (\ref{coupleddiracequations}), we obtain $-\ddot{\alpha}_{R,L}+U_{R,L}(z)\alpha_{R,L}=m^2\alpha_{R,L}$, where 
\begin{equation}
U_{R,L}= e^{-2A}(\epsilon_1 h^p \sqrt{R}+\epsilon_2 e^{A}
\dot{h} )^2\pm \frac{d}{dz}[e^{-A}(\epsilon_1 h^p \sqrt{R}+\epsilon_2 e^{A}
\dot{h} )].
\end{equation}
The superpotential $W=e^{-A}(\epsilon_1 h^p \sqrt{R}+\epsilon_2 e^{A}
\dot{h} )$ has the expression
\begin{eqnarray}
W(z)&=&\epsilon_1 \frac{\sqrt{2}\lambda}{\sqrt{1+\lambda^2 z^2}} tg^{-1}\left(\frac{\lambda z}{\sqrt{1+\lambda^2 z^2}}\right)\left(\frac{5}{3}+\frac{1}{1+\lambda^2 z^2}\right)\nonumber\\
&+& \epsilon_2 \frac{2\lambda}{\sqrt{1+\lambda^2 z^2}(1+2\lambda^2 z^2)}.
\end{eqnarray}
Note that for $p=1$ the first term is odd (as usual in thick braneworld models), whereas the second term is even. The respective potential has the SUSY form
$U_{R,L}=W^2 \pm \dot{W}$. For $p=1$, we plotted the superpotential (fig.\ref{superpotential}) and the corresponding left potential $U_L$ (fig.\ref{leftpotentialdisplay}) for some values of $\epsilon_1$ and $\epsilon_2$. For $\epsilon_2 =0$, the superpotential is an odd function of $z$ and the left-handed potential has the usual volcano shape. For $\epsilon_2 \neq 0$, both $W$ and $U_L$ become asymmetrical.


\begin{figure}[htb] 
      \begin{minipage}[b]{0.5 \linewidth}
          \includegraphics[width=\linewidth]{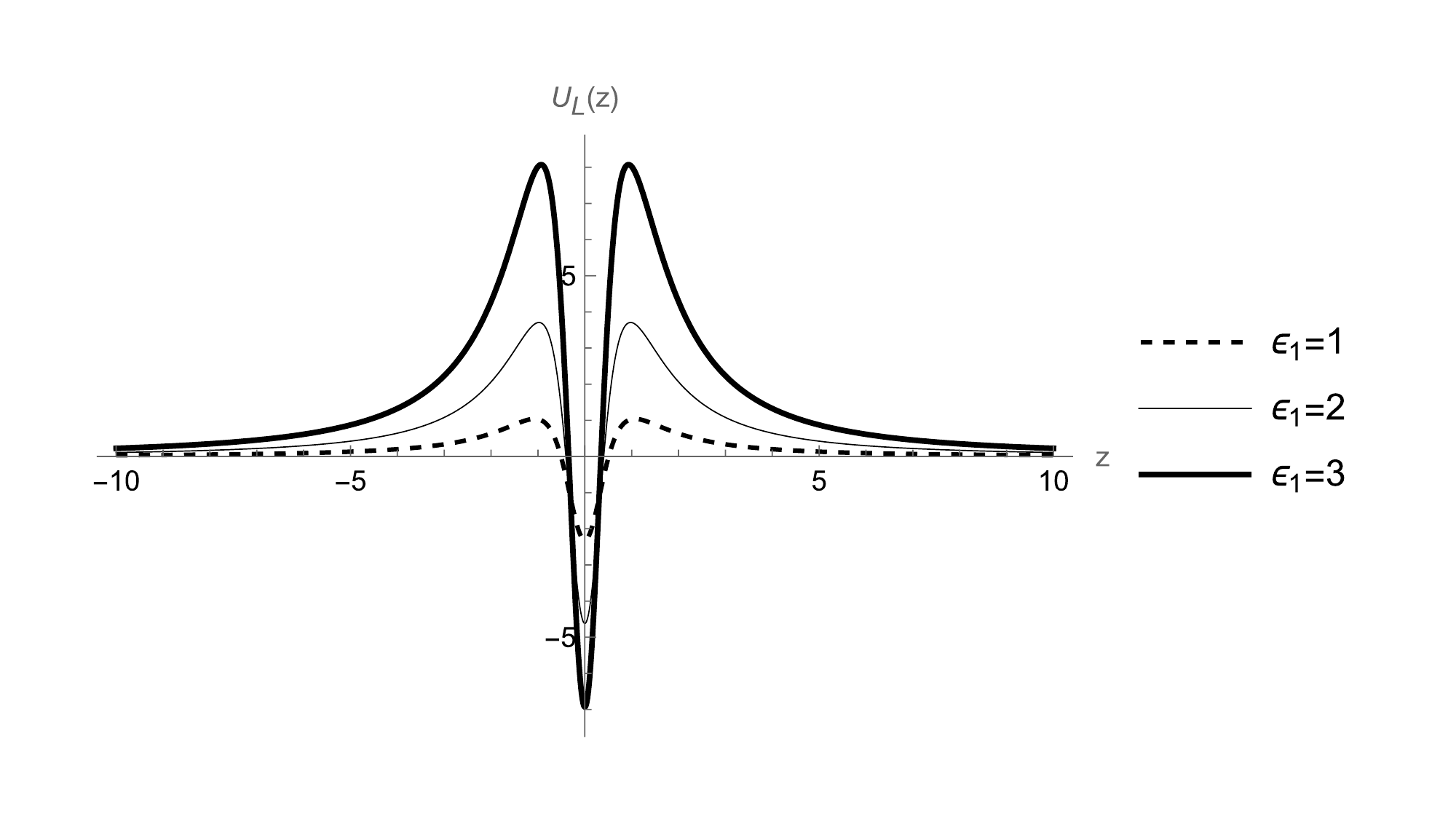}\\
           \caption{Left-handed potential $U_L$ for $\epsilon_2 =0$, $\lambda=p=1$.}
          \label{lefthandedp}
       \end{minipage}\hfill
       \begin{minipage}[b]{0.5\linewidth}
           \includegraphics[width=\linewidth]{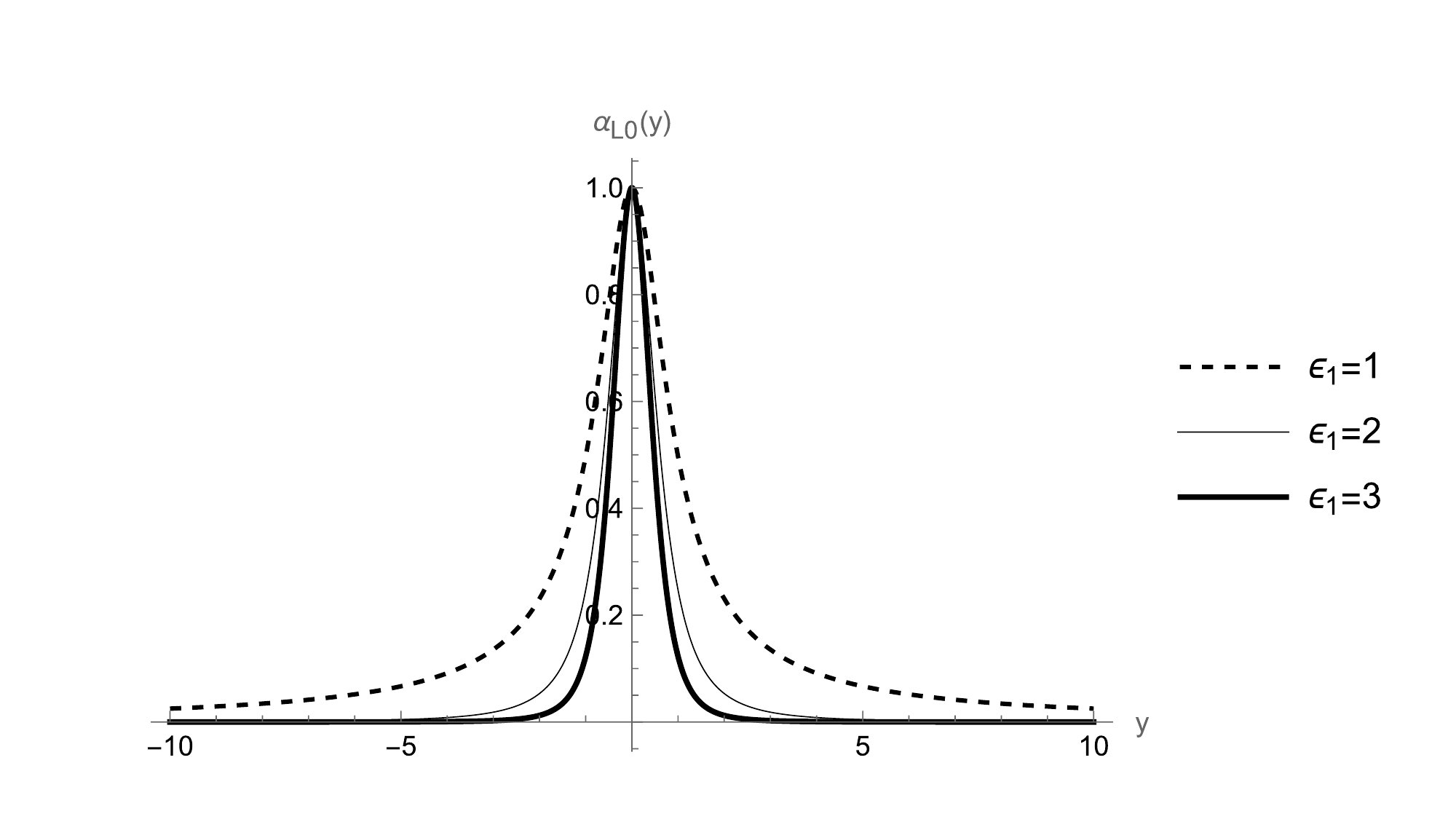}\\
           \caption{Left-handed massless mode for $\epsilon_2 =0$, $\lambda=p=1$..}
          \label{fermionicmassless}
       \end{minipage}
\end{figure}

By fixing $\epsilon_2 =0$ and varying $\epsilon_1$, we observe by the fig.(\ref{lefthandedp}) and fig.(\ref{fermionicmassless}) that the massless lefthanded fermionic mode gets more localized around the origin as $\epsilon_1$ grows. Therefore, the geometric non-minimal coupling with $\epsilon_2=0$ presents an alternative to the well-known Yukawa coupling \cite{casa}.

\section{Final remarks and perspectives} \label{section6}
In this work we proposed torsional 5D braneworld scenarios whose dynamics is equivalent to a tensor-vector gravity theory. The torsion steams from a background vector in the so-called Lyra geometry.

Inasmuch as the tensor vector is covariantly constant, the squared norm of this Lyra vector can be regarded as a source for the bulk cosmological constant. Assuming the Lyra vector as the derivative of a scalar field we obtained an isotropic scenario similar to the coupled two fields Bloch brane \cite{blochbrane}. 

The background vector allow us to define non-minimal couplings to the bulk fields. For a scalar field in the isotropic model a curvature mass-like term modifies the interaction between the KK modes and the brane. Positive non-minimal coupling results in a symmetric potential barrier at the origin, and a negative one leads to a potential well. In fact, the non-minimal coupling breaks the analogue SUSY symmetry.
Thus, the chances to detect massive modes on the brane is reduced in this scenario. Moreover, the spacelike Lyra vector breaks the $Z_2$ symmetry trying to localize the massless mode out of the brane. For the Abelian gauge vector we proposed a modified kinetic term with a birefringence-like factor proportional to the norm of the Lyra vector. If $b_M=\partial_M h$, this coupling resembles a derivative dilaton-gauge vector interaction \cite{Kehagias2001}. 

The spinor connection is modified by the torsion that yields a derivative $\bar{\Psi}i\Gamma^M \partial_M h \Psi$ coupling. In addition, we proposed a non-minimal coupling of form $hR\bar{\Psi}\Psi$, which can be considered as a pure-geometric Yukawa-like coupling. It turns out that only the non-minimal coupling leads to relevant modifications to the fermionic KK modes. For a vanishing coupling constant $\epsilon_2$, the left-handed potential has the usual volcano shape and the he massless left handed  fermionic  mode  gets  more  localized  around  the  origin  as $\epsilon_1$ grows.This results leads to the conclusion that the geometric non-minimal coupling with $\epsilon_2=0$ presents an alternative to the well-known Yukawa coupling \cite{casa}.
It is worthwhile to mention that the KK reduction provides a CPT-odd SME Lorentz violating term $i\bar{\psi}\gamma^\mu b_\mu \psi$ \cite{kostelecky1}. We argue that by considering more non-minimal couplings other SME terms could be generated and their experimental test could be used to obtain upper bound to the bulk-brane torsion. 

The close relationship between Lyra and Weyl geometries in braneworld models is another topic which was addressed. In a integrable Weyl geometry and using a general conformal invariant action, some pure-geometric thick branes were found \cite{weyl1, weyl2}. Therefore, the investigation of a pure-geometric thick branes in a conformal Lyra geometry, and the effects upon the fields thereof, it is a worth extension of the present work.

\section{Acknowledgments}
\hspace{0.5cm}The authors thank the Funda\c{c}\~{a}o Cearense de Apoio ao Desenvolvimento Cient\'{\i}fico e Tecnol\'{o}gico (FUNCAP)(PNE-0112-00061.01.00/16), and the Conselho Nacional de Desenvolvimento Cient\'{\i}fico e Tecnol\'{o}gico (CNPq) for grants 308638/2015-8 (CASA).

\end{document}